\pdfoutput=1
\documentclass[reprint,aps,prx,showpacs,twocolumn]{revtex4-1}

\usepackage{natbib}
\usepackage{graphicx}
\usepackage[utf8]{inputenc}
\usepackage{mathptmx}
\usepackage{epsfig,amsopn}
\usepackage{graphicx}
\usepackage{amsmath,amssymb}
\usepackage{natbib,color}
\usepackage{braket}
\usepackage{float}
\usepackage{enumerate}
\usepackage[normalem]{ulem}
\usepackage{todonotes}
\allowdisplaybreaks[1]

\newcommand{\eref}[1]{Eq.~(\ref{#1})}%
\newcommand{\fref}[1]{Fig.~\ref{#1}} %
\newcommand{\sgn}[1]{\mathrm{sgn}({#1})}%

\def\bea{\begin{eqnarray}}
\def\eea{\end{eqnarray}}

\DeclareMathOperator{\sign}{sgn}

\begin{document}

\title{Invariants of motion with stochastic resetting and space-time coupled returns}

\author{Arnab Pal$^{1}$}
\email{arnabpal@mail.tau.ac.il}

\author{\L{}ukasz Ku\'smierz$^{2}$}
\email{nalewkoz@gmail.com}

\author{Shlomi Reuveni$^{1}$}
\email{shlomire@tauex.tau.ac.il}

\affiliation{\noindent \textit{$^{1}$School of Chemistry, The Center for Physics and Chemistry of Living Systems, The Raymond and Beverly Sackler Center for Computational Molecular and Materials Science,\\ \& The Mark Ratner Institute for Single Molecule Chemistry, Tel Aviv University, Tel Aviv 6997801, Israel}}

\affiliation{\noindent \textit{$^{2}$Laboratory for Neural Computation and Adaptation, RIKEN Center for Brain Science, 2-1 Hirosawa, Wako, Saitama 351-0198, Japan}}

\date{\today}

\begin{abstract}
    Motion under stochastic resetting serves to model a myriad of processes in physics and beyond, but in most cases studied to date resetting to the origin was assumed to take zero time or a time decoupled from the spatial position at the resetting moment. However, in our world, getting from one place to another always takes time and places that are further away take more time to be reached. We thus set off to extend the theory of stochastic resetting such that it would account for this inherent spatio-temporal coupling. We consider a particle that starts at the origin and follows a certain law of stochastic motion until it is interrupted at some random time. The particle then returns to the origin via a prescribed protocol. We study this model and surprisingly discover that the shape of the steady-state distribution which governs the stochastic motion phase does not depend on the return protocol. This shape invariance then gives rise to a simple, and generic, recipe for the computation of the full steady-state distribution. Several case studies are analyzed and a class of processes whose steady-state is completely invariant with respect to  the speed of return is highlighted. For processes in this class we recover the same steady-state obtained for resetting with instantaneous returns---irrespective of whether the actual return speed is high or low. Our work significantly extends previous results on motion with stochastic resetting and is expected to find various applications in statistical, chemical, and biological physics.
    
\end{abstract}

\maketitle

\section{Introduction}
Stochastic motion with stochastic resetting is of considerable interest due to its broad applicability in statistical \cite{Restart1,Restart2,KM,restart_conc3,restart_conc2}, chemical \cite{Restart-Biophysics1,Restart-Biophysics4,Restart-Biophysics5,Restart-Biophysics6,Restart-Biophysics2}, and biological physics \cite{Restart-Biophysics3,Restart-Biophysics8}; and due to its importance in computer science \cite{Luby,Gomes,Montanari,Steiger} and the theory of search and first-passage \cite{Restart-Search1,Chechkin,Restart-Biophysics7}. Particularly, in statistical physics, such motion has become a focal point of recent studies owing to the rich non-equilibrium \cite{Restart1,Restart2,KM,restart_conc3,restart_conc2,restart_conc5,Satya-refractory,thermo} and first-passage  \cite{ReuveniPRL,PalReuveniPRL,branching_I,branching_II,Landau,Belan} phenomena it displays. 

Motion with stochastic resetting is fairly simple to understand: a process on the run is interrupted at a random point in time and consequently reset to start anew. Noteworthy in this regard is the paradigmatic, Evans-Majumdar, model for diffusion with stochastic resetting  \cite{Restart1,Restart2}. This model has led to a large volume of work covering diffusion with resetting in the presence of a potential field \cite{restart_conc2,Ray,Ahmad}, in different geometrical confinements \cite{Christou,restart_conc8,localtimer,restart_conc9}, in higher dimensions \cite{restart_conc6}, with  non-Poissonian resetting protocols \cite{Restart-Search3,restart_conc18,kusmierz2018robust,Restart4,Restart5}, with interactions \cite{restart_conc1,restart_conc12,SEP}, and more. The model was further extended to study other, i.e., non-diffusive, stochastic processes under resetting \cite{restart_conc7,Bodrova1,Bodrova2,restart_conc16,Restart3,restart_conc21,subCTRW,Restart-Search1,Restart-Search2,Satya-RT,telegraphic,transport1}. 

\begin{figure}[t]
\centering
\includegraphics[width=\linewidth]{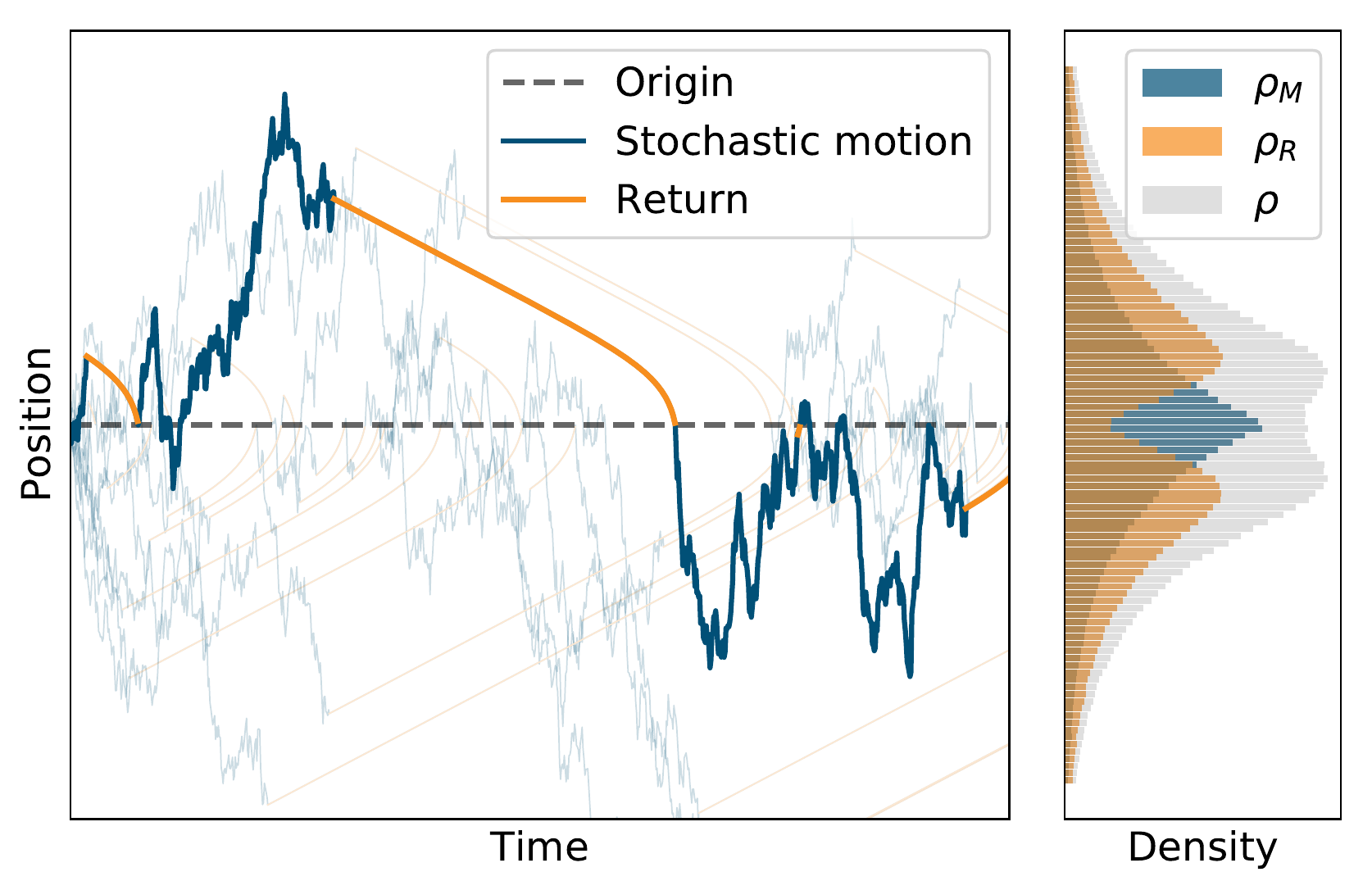}
\caption{An illustration of motion with stochastic resetting and space-time coupled returns to the origin. The dynamics consists of two phases: (i) stochastic motion (blue), wherein the particle moves according to a given law [in the example: Brownian motion], and (ii) return (orange), wherein upon resetting the particle returns to its initial position according to some protocol [in the example with speed $v_r(x) = 0.5 + 5 \exp(-5 |x|)$]. We study the probability densities $\rho_M$ (motion), $\rho_R$ (return), and $\rho=\rho_M+\rho_R$ (total) for this wide and diverse class of processes; and develop a simple, invariance based, recipe that allows their computation at the steady-state.}
\label{fig:sample-traj}
\end{figure}

In the Evans-Majumdar model, and many of its extensions, resetting is taken to be instantaneous. This is quite unrealistic as it means that upon resetting the diffusing particle returns to its initial position with an infinite velocity. However, in reality, a particle cannot return (or be returned) to the origin in zero time. Several attempts were made to address this issue e.g., by incorporating an overhead time (refractory period) that follows each resetting event \cite{Restart-Biophysics1,Restart-Biophysics2,Ahmad,transport1,Satya-refractory}; but in all these attempts it was assumed that there is no direct coupling between the overhead time and the position of the particle at the resetting moment---which is again non-physical since returning from afar usually takes longer. To address this point, we have recently introduced a comprehensive theory for first-passage under space-time coupled resetting, a.k.a, home-range search, which does not make any assumptions on the underlying stochastic motion and is furthermore suited to treat generic return and home-stay strategies \cite{HRS}. In this paper, we set aside first-passage questions in attempt to understand spatial properties of motion with stochastic resetting and space-time coupled returns to the origin (Fig. \ref{fig:sample-traj}). 

\section{Markov processes with stochastic resetting and space time coupled returns}
\label{SecII}
We start with a Markovian setup which we later on generalize. Consider a particle undergoing stochastic motion and further assume that the propagator which describes this stochastic process obeys the following Master equation
\begin{equation}
    \partial_t \rho(x,t) = 
    \mathcal{L}_0 \rho(x,t),
    \label{eq:balance-dynamics-active-main}
\end{equation}
where $\mathcal{L}_0$ is the infinitesimal generator of the process without resetting. To introduce  stochastic resetting with instantaneous returns into the model imagine that at any small time interval $\Delta t$ the particle's motion can be reset with probability $r\Delta t$. If such resetting happens, the particle will teleport back to the origin and start its motion anew. The corresponding master equation then reads  
\bea
\partial_t\rho(x,t)=\mathcal{L}_0 \rho(x,t)-r \rho(x,t)+r \delta(x)~.
\label{inst-propagator}
\eea

In this section, we will construct a set of master equations, akin to \eref{inst-propagator}, to describe motion  with stochastic resetting and space-time coupled returns to the origin. We  consider a situation in which the particle returns to the origin with a space dependent velocity ${v(x) = -\sign(x)v_r(x)}$, where $v_r(x)>0$ is the return speed and $\sign(x)$ is the  signum function (takes the value: $+1$ if $x>0$, $-1$ if $x<0$, and zero otherwise). Note that the signum function appears here because returning particles move in the direction of the origin, i.e., to the left if $x>0$ and to the right if $x<0$. In what follows we will assume that the return speed is continuous in the vicinity of the origin, i.e. $v_r(0) = \lim_{x\to 0} v_r(x)$. 

Similar to the above, we will  denote the propagator of our process by $\rho(x,t)$, but discriminate between two different phases of motion: (i) the stochastic motion phase in which the particle performs stochastic motion according to the law described in \eref{eq:balance-dynamics-active-main}; and (ii) the return phase in which, upon resetting, the particle returns to its initial position as described above. Our propagator thus has two contributions, one from each phase, and it can be written as
\bea
\rho(x,t)=\rho_M(x,t)+\rho_R(x,t)~,
\eea
where $\rho_M(x,t)$ and $\rho_R(x,t)$ correspond to the probability densities governing the stochastic motion and return phases respectively. It is clear that $\rho_M(x,t)$ and $\rho_R(x,t)$ are not individually normalized as their sum is the total probability density $\rho(x,t)$ which is normalized to one. Evidently the probabilities to find the particle in the motion and return phases are given by
\begin{align}
\begin{split}
    p_M(t) \equiv \text{Prob}(\text{\textit{motion}}) &= \int\limits_{-\infty}^{\infty} \mathrm{d}y~\rho_M(y,t), 
    \\
    p_R(t) \equiv \text{Prob}(\text{\textit{return}}) &= \int\limits_{-\infty}^{\infty} \mathrm{d}y~\rho_R(y,t),
    \label{pa-pb}
\end{split}
\end{align}
where $p_M(t)+p_R(t)=1$ at all times. 

We now set to find equations for $\rho_M(x,t)$ and $\rho_R(x,t)$, and thus for the propagator $\rho(x,t)$ which describes our process. We start by considering the time evolution of the position distribution in the return phase $\rho_R(x,t)$. To this end, we recall that particles in the return phase move at velocity ${v(x) = -\sign(x)v_r(x)}$. The probability flux at $x$ due to such particles is thus $\partial_x[\sgn{x}v_r(x)\rho_R(x,t)]$. In addition, we note that particles enter the return phase from the stochastic motion phase at a rate $r$, and that the probability flux at $x$ due to such particles is $r\rho_M(x,t)$. Summing over the two possibilities above gives 
\bea
    \partial_t \rho_R(x,t) &=&
    \partial_x[\sgn{x}v_r(x)\rho_R(x,t)]+ r\rho_M(x,t) 
     \nonumber
    \\
      \nonumber
    \\
    &-&
    2\delta(x)v_r(0)\rho_R(0,t)~, 
    \label{return-delta}
\eea
where the last term on the right hand side serves as a sink and accounts for the fact that returning particles switch to stochastic motion mode upon arrival to the origin. Finally, we observe that taking the spatial derivative on the right side of \eref{return-delta} cancels the last term and leaves us with
\begin{equation}
    \partial_t \rho_R(x,t) = 
    \sign{(x)}  \partial_x\left[ v_r(x)  \rho_R(x,t)\right] + r \rho_M(x,t)~.
    \label{eq:balance-dynamics-zombie-main}
\end{equation}

We now turn our attention to the time evolution of the position distribution in the stochastic motion phase $\rho_M(x,t)$. To this end, we observe that a stochastically moving particle will be found at position $x$ at time $t+\Delta t$ if at time $t$ it was positioned at $x-\Delta x$ and provided that in the following time interval, $\Delta t$, it moved an increment $\Delta x$. Noting that the probability to stay in the stochastic motion phase within this latter time interval is $(1-r\Delta t)$ we have 
\bea
\rho_M(x,t+\Delta t) &=& (1-r\Delta t) \langle \rho_M(x-\Delta x,t)   \rangle \nonumber \\ &+&\delta(x)~\int\limits_{-v_r(0^-) \Delta t}^{ v_r(0^+) \Delta t}~\mathrm{d}z~\rho_R(z,t) 
+ O(\Delta t^2),
\label{eq:balance-dynamics-active-main1}
\eea
where the average in the first term on the right hand side is taken with respect to the random increment $\Delta x$, and the second term acts as a source which accounts for the inflow at the origin due to particles returning from the domain $\left[ -v_r(0^-) \Delta t,v_r(0^+) \Delta t  \right]$ and consequently switching to stochastic motion  mode. Taking the $\Delta t\to 0$ limit in \eref{eq:balance-dynamics-active-main1}, the corresponding continuous-time evolution equation reads
\bea
    \partial_t \rho_M(x,t) = \mathcal{L}_0 \rho_M(x,t)-r \rho_M(x,t)  + 2 \delta(x) {v}_r(0) \rho_R(0,t),\hspace{0.3cm}
    \label{eq:balance-dynamics-active-main2}
\eea
where we assumed, without much loss of generality, that $\rho_R(x,t)$ is continuous around ${x=0}$. Equations (\ref{eq:balance-dynamics-zombie-main}) and (\ref{eq:balance-dynamics-active-main2}) constitute a set of two coupled partial differential equations that should be solved if one would like to obtain a full, time dependent, description of stochastic motion with resetting and space time coupled returns to the origin. A detailed account on how this can be done for simple Brownian motion is given in \cite{invariance}, and while results there can extended, we hereby focus our attention at the steady-state. 

\section{Steady-State: Markovian Setting}
\label{SecIII}

At the steady-state \eref{eq:balance-dynamics-zombie-main} reduces to
\begin{equation}
\sign{(x)} \partial_x  \left[ v_r(x) \rho_R(x) \right] 
=  - r \rho_M(x),
\label{eq:passive-from-active-derivative-main}
\end{equation}
with $\rho_M(x)$ and $\rho_R(x)$ standing respectively for the stationary distributions describing the stochastic motion and return phases. Taking the stationary limit in \eref{eq:balance-dynamics-active-main2}, we also have 
\bea
\mathcal{L}_0 \rho_M(x)-r \rho_M(x)
    +2 \delta(x) {v}_r(0) \rho_R(0)=0~.
    \label{ss-search}
\eea
A close look at \eref{ss-search} suggests that in order to compute $\rho_M(x)$ one first needs to find $\rho_R(0)$. Integrating \eref{eq:passive-from-active-derivative-main} over the real line, we find
\bea
\rho_R(0)=\frac{rp_M}{2{v}_r(0)},
\label{eq:rhoR-origin-from-vr}
\eea
where we have imposed the natural boundary conditions $\underset{x \to \pm \infty}{\mathrm{lim}} v_r(x)\rho_R(x)=0$ 
and defined $p_M$ to be the steady-state probability to find the particle at the stochastic motion phase. Substituting this result back into in \eref{ss-search} one finds
\bea
\mathcal{L}_0 \rho_M(x)-r \rho_M(x)
    +rp_M \delta(x)=0 \label{ss-search1}~.
\eea
We now note that $\rho_M(x)$ can also be written as  
\begin{equation} 
\rho_M(x) = p_M \rho(x|\text{\textit{motion}}),
\label{eq:active-fast-same-main}
\end{equation}
where $\rho(x|\text{\textit{motion}})$ is the conditional probability density to find the particle at $x$ given that it is in the stochastic motion phase. Dividing both sides of \eref{ss-search1} by $p_M$, we see that at the steady-state 
\bea
\mathcal{L}_0 \rho(x|\text{\textit{motion}})-r \rho(x|\text{\textit{motion}})
    +r\delta(x)=0~, \label{ss-search2}
\eea
which is identical to the stationary limit of \eref{inst-propagator}. Thus the steady-state of $\rho(x|\text{\textit{motion}})$ in our model is identical to that which is obtained for the total density in a model where returns to the origin are instantaneous (limit of $v_r(x) \to \infty$). Namely, letting $\rho_{\mathrm{\infty}}(x)$ stand for the steady-state solution of \eref{inst-propagator}, we have \bea
\rho(x|\text{\textit{motion}})=\rho_{\mathrm{\infty}}(x)~.
\label{conditional}
\eea
Concluding, we see that the shape of the steady-state density describing the stochastic motion phase is completely invariant to the profile of the return speed $v_r(x)$ which is the first invariance result we establish in this paper. We will now show that the result in \eref{conditional} is extremely general and that it remains valid even beyond the Markovian setup we have considered so far. We will utilize this fact to provide a simple, and general, recipe for the computation of steady-state distributions in our model.

\section{Steady-state: General Setting}
\label{SecIV}

To prove \eref{conditional} in a general setting we recall that  probability densities do not evolve with time at the steady-state. In particular, the probabilities to be in the stochastic motion and return phases are constant. This in turn means that the probability flux from the stochastic motion phase to the return phase, due to resetting with rate $r$, must be exactly balanced by an opposing probability flux. However, the only place where returning particles switch back into stochastic motion is at the origin. Taking the perspective the stochastic motion phase, we see that at steady-state the outgoing probability flux is instantaneously balanced by an incoming probability flux that emanates at the origin. Since the exact same thing happens at the steady-state of a model where returns to the origin are instantaneous, and since the dynamics of stochastic motion in the bulk is the same regardless of the return protocol, \eref{conditional} must hold in general.     

The invariance described by \eref{conditional} allows us to rewrite \eref{eq:active-fast-same-main} in the following form
\begin{equation} 
\rho_M(x) = p_M \rho_{\mathrm{\infty}}(x)~.
\label{eq:active-fast-same-main-2}
\end{equation}
Moreover, noting that the derivation of  \eref{eq:passive-from-active-derivative-main} did not assume that the underlying stochastic process is Markovian, we solve it to obtain 
\begin{equation}
\rho_R(x) 
=\frac{rp_M}{v_r(x)}
\sign(x)
\int\limits_x^{\sign(x)\infty}
\mathrm{d} y
~ \rho_{\mathrm{\infty}}(y)~,
\label{eq:passive-from-active-main}
\end{equation}
where we have again imposed $\underset{x \to \pm \infty}{\mathrm{lim}} v_r(x)\rho_R(x)=0$. We note in passing that \eref{eq:passive-from-active-main} remains valid even if $v_r(x)$ or $\rho_\infty(x)$ have a discontinuity at $x=0$, albeit the fact that one then needs to separate treatment for the positive and negative branches of the x-axis. 

To find $p_M$ in the above equations, we observe that this probability is identical to the time fraction the particle spends in stochastic motion at the steady-state. The mean time spent at the stochastic motion phase is $1/r$ (inverse of restart rate). On the other hand, the time spent returning from position $x$ is 
\bea
\tau(x)=\sgn{x}\int^{x}_{0}~\frac{\mathrm{d}z}{v_r(z)}, 
\eea
which means that the mean time spent at the return phase is 
\bea
\langle \tau(x) \rangle= \int\limits_{-\infty}^{\infty}~\mathrm{d}x~ \tau(x) \rho(x|\text{\textit{motion}})=\int\limits_{-\infty}^{\infty}~\mathrm{d}x~ \tau(x)~\rho_{\infty}(x)~.\hspace{0.3cm}
\label{taux}
\eea
Utilizing this we have
\bea
p_M=\frac{1/r}{1/r+\langle \tau(x) \rangle}~.
\label{pa-exact}
\eea
Finally, we note that another way to find $p_M$ is by utilizing the fact that the total probability density $\rho(x)=\rho_R(x)+\rho_M(x)$ is normalized to one.

Equations (\ref{eq:active-fast-same-main-2})-(\ref{pa-exact}) provide a simple recipe for the evaluation of steady-state distributions governing motion with stochastic resetting and space-time coupled returns. Evaluation is done in terms of $\rho_{\mathrm{\infty}}(x)$ which is the steady-state distribution obtained for the case of instantaneous returns. Evaluating $\rho_{\mathrm{\infty}}(x)$ itself is now common practice as it can be linked to the Laplace transform of the propagator, $\rho_{r=0}(x,t)$, that governs stochastic motion in the absence of resetting through the renewal formalism 
\bea
\rho_{\mathrm{\infty}}(x)=\int_0^\infty~\mathrm{d}t~re^{-rt}~\rho_{r=0}(x,t)=r\tilde{\rho}_{r=0}(x,r)~.
\label{renewal-inst}
\eea
Concrete examples that illustrate how the above procedure can be applied in practice are considered below.

\section{Examples}

In this section, we will present a series of exactly solvable case studies to demonstrate the power of our approach. We start with the case of diffusion. 

\subsection{Diffusion}

Consider our model for simple diffusion with a diffusion coefficient $D$, stochastic resetting rate $r$, and a constant return speed $v_r$. We have recently shown that the total density, and the densities of the diffusive and return phases, can be computed for this case study at all times by solving  \eref{eq:balance-dynamics-zombie-main} and \eref{eq:balance-dynamics-active-main2} (see \cite{invariance} for details). In particular, the  steady-state solution can be obtained in this way, but in this subsection we will take an alternative approach and utilize the formalism prescribed above to get the same result a bit more directly.  

First, recall that the steady-state density for diffusion with stochastic resetting and instantaneous returns is known and can be readily obtained by plugging in the Gaussian propagator of simple diffusion into \eref{renewal-inst}. This gives 
\bea
\rho_{\mathrm{\infty}}(x)=\frac{\alpha_0}{2}~e^{-\alpha_0|x|}~,
\label{DUR_in}
\eea
where $\alpha_0=\sqrt{\frac{r}{D}}$ can be interpreted as the inverse of the average distance traveled by the particle between two resetting events \cite{Restart1}. Substituting this result into \eref{taux} and utilizing \eref{pa-exact}, we obtain
\bea
p_M=\left( 1+\frac{r}{v_r \alpha_0}  \right)^{-1}~,
\eea
which by use of \eref{eq:active-fast-same-main-2} gives the density in the stochastic motion, i.e., diffusive, phase 
\bea
\rho_M(x)=p_M\rho_{\mathrm{\infty}}(x)=\frac{\alpha_0p_M}{2}e^{-\alpha_0|x|}~.
\eea
As expected, this density identifies with $\rho_{\mathrm{\infty}}(x)$ up to the scaling factor $p_M$. The density in the return phase can be computed using \eref{eq:passive-from-active-main} and we find
\bea
\rho_R(x)=\frac{rp_M}{2v_r} e^{-\alpha_0|x|}~,
\eea
which once again identifies with $\rho_{\mathrm{\infty}}(x)$ up to a scaling factor.  

Finally, the total density can be obtained by summing over $\rho_M(x)$ and $\rho_R(x)$ to give
\bea
\rho(x)&=&\frac{p_M}{2} \left( \alpha_0+\frac{r}{v_r} \right)  e^{-\alpha_0|x|} =\frac{\alpha_0}{2}e^{-\alpha_0|x|}~.
\eea
Interestingly, this form is completely invariant to the return speed $v_r$ and therefore identical to the result obtained for instantaneous returns [\eref{DUR_in}]. Surprisingly, one can moreover show that for diffusion this invariance extends beyond the steady-state, i.e.,  $\rho(x,t)$=$\rho_\infty(x,t)$ for any finite time $t\geq0$, and we refer the reader to \cite{invariance} for details.

\begin{figure*}[t]
\centering
\includegraphics[width=\linewidth]{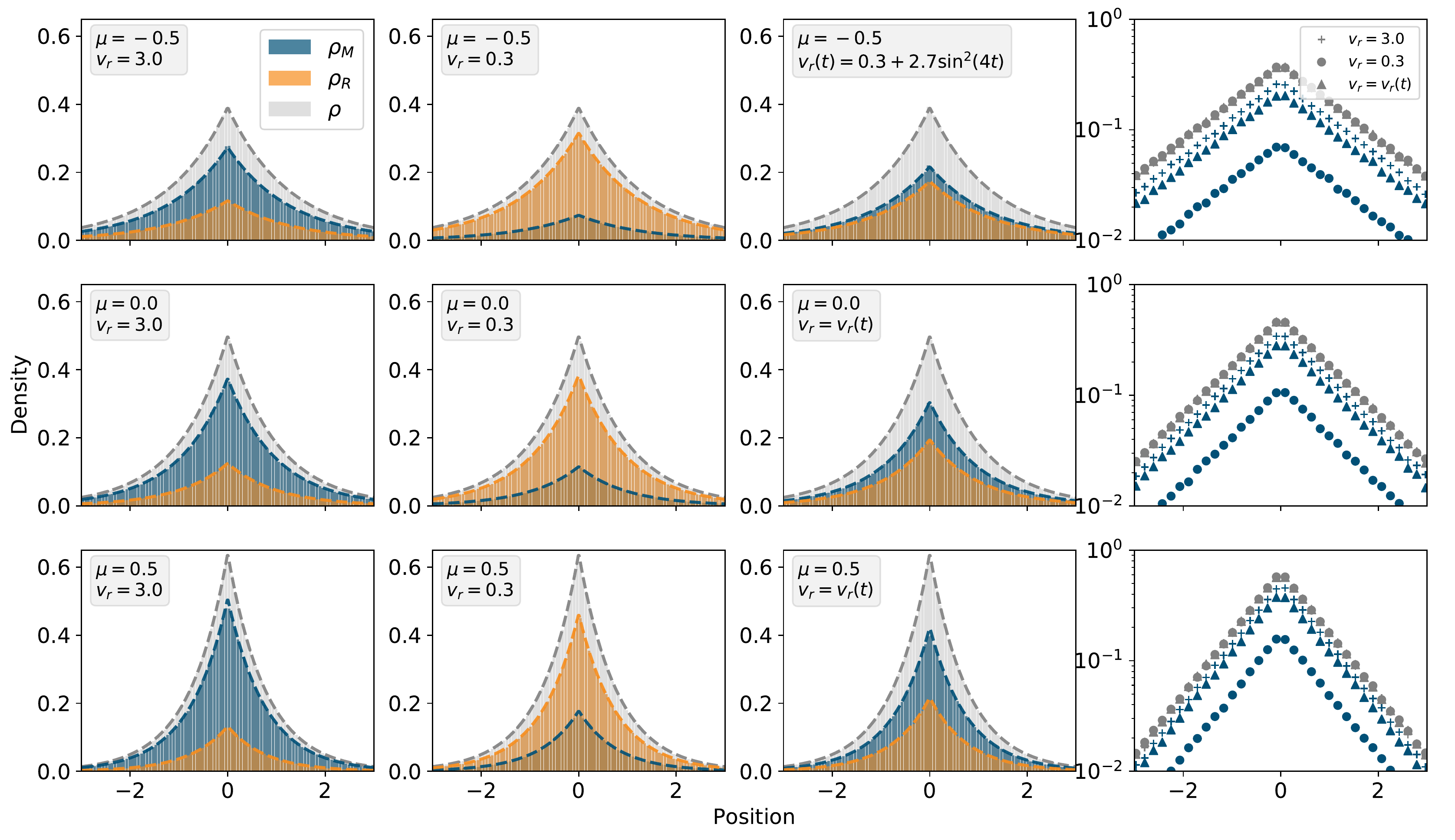}
\caption{Theory and numerical simulations for diffusion in  $V$-shaped potential, $V(x)=\mu|x|$, with stochastic resetting and space-time coupled returns to the origin. Each of the first three columns represents a different return protocol. In the first two columns the return speed was correspondingly set to $v_r=3$ and $v_r=0.3$; and in the third column the return speed was allowed to oscillate with time such that $v_r(t)=0.3+2.7 \sin^2(4 t)$ with $t$ being the time elapsed since the last resetting epoch. Rows correspond to different values of the parameter $\mu$ which governs the strength of the potential. Histograms represent results obtained via computer simulations and dashed lines correspond to theoretical predictions with the exception of the blue and orange lines in the third column. There, in the absence of a closed form formula for $p_M$, we estimated the probability to be in the diffusive phase from simulations. In all cases, we took $D=1$, $r=1$, initiated the process at the origin, and integrated numerically with $\Delta = 0.01$ to generate $10^6$ independent samples of the particle's position at  final time $T=10$. Excellent agreement with theory is found. In the last column, simulation results for the total steady-state density $\rho(x)$ (gray), and the density at the diffusive phase $\rho_M(x)$ (blue), are plotted on a semi-log scale for the different cases. It can be seen that $\rho_M(x)$ identifies with $\rho(x)$ up to a scaling factor as predicted. One can also observe that the total density $\rho(x)$ is completely invariant to the return speed as predicted by \eref{Inv-V}.}
\label{fig:hist-success}
\end{figure*}

\subsection{Diffusion in $V$-shaped potential}

Simple diffusion is not the only process whose steady-state distribution under stochastic resetting is invariant to the return speed. Indeed, one can easily convince himself that this would also be the case for any stochastic process whose steady-state distribution under stochastic resetting with instantaneous returns is Laplace, i.e., has the form prescribed in \eref{DUR_in}. For example, consider diffusion in the presence of a V-shaped potential $V(x)=\mu|x|$. The steady-state of this process under stochastic resetting and instantaneous returns is known and can be computed by solving the steady-state limit of \eref{inst-propagator} with the proper infinitesimal generator $\mathcal{L}_0$ \cite{restart_conc2}. This gives 
\bea
\rho_{\mathrm{\infty}}(x)=\frac{\alpha_{\mu}}{2}~e^{-\alpha_{\mu}|x|}~,
\label{DURV_in}
\eea
where $\alpha_{\mu}=\frac{\mu+\sqrt{\mu^2+4Dr}}{2D}$. Comparing \eref{DURV_in} with \eref{DUR_in}, we see that  all the results in the previous subsection carry through with $\alpha_{\mu}$ replacing $\alpha_{0}$. In particular, the total density is given by 
\bea
\rho(x)&=& \frac{\alpha_{\mu}}{2}e^{-\alpha_{\mu}|x|}~,
\label{Inv-V}
\eea
which is again completely invariant to the return speed. The results for diffusion in $V$-shaped potential are corroborated against numerical simulations in  Fig.~\ref{fig:hist-success}, which in addition reveals that the invariance displayed by \eref{Inv-V} continues to hold even in situations where the return speed depends on the time that elapsed since the last resetting epoch. 

\subsection{Telegraph process}
\begin{figure}[t]
\centering
\includegraphics[width=\linewidth]{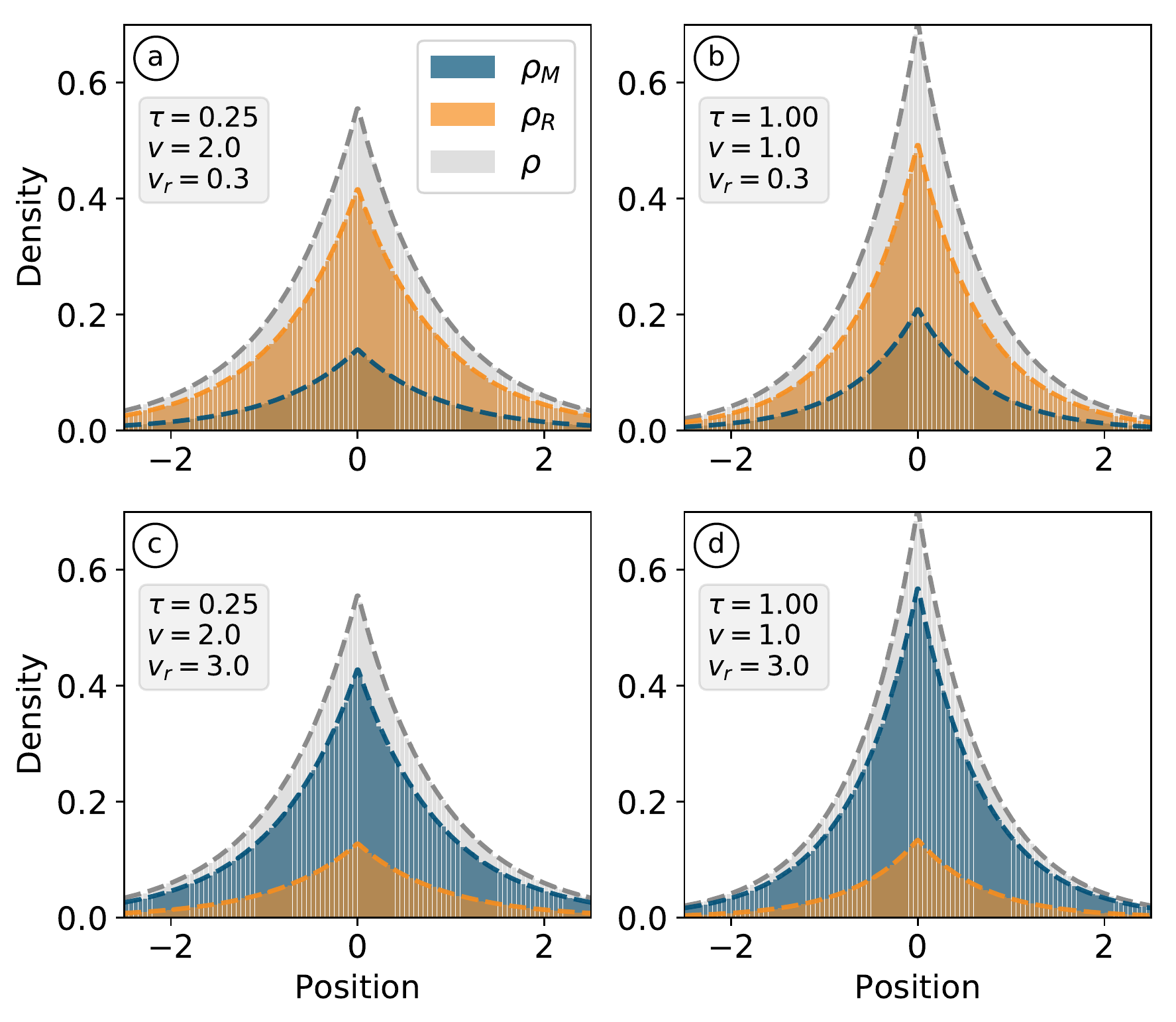}
\caption{
Theory and numerical simulations for a telegraph process with stochastic resetting and constant speed returns to the origin. Plots are made for four different combinations of the mean switching time $\tau$, the ballistic motion velocity $v$, and the return velocity $v_r$. Histograms represent results obtained via computer simulations, the corresponding dashed lines our theoretical predictions. In all cases, we took $r=1$, initiated the process at the origin, and integrated numerically with a time step $\Delta = 0.01$ to generate $10^6$ independent samples of the particle's position at a final time $T=20$. Excellent agreement with theory is found.}
\label{fig:hist-tele}
\end{figure}

As we have discussed in section \ref{SecIV}, the results we have obtained are not limited to Markovian processes whose propagator obeys \eref{eq:balance-dynamics-active-main}. To demonstrate this, we consider a one dimensional telegraph process in which a particle switches stochastically between ballistic motion with a positive velocity $+v$ to ballistic motion with a negative velocity $-v$. As a result, the duration $t$ and run length $|x|$ of each ballistic motion session are coupled via: $|x|=vt$. Consequently, the joint distribution for the session duration and displacement is given by  $\psi(x,t)=\frac{1}{2}\delta(|x|-vt)\phi(t)$ with $\phi(t)$ standing for the distribution of the stochastic switching time between the positive and negative modes of motion. In particular, when this is governed by the exponential distribution, $\phi(t)=\tau^{-1}e^{-t/\tau}$ $(t\geq0)$, and in the absence of resetting, the propagator of the telegraph process is known to have the following form in Fourier-Laplace space \cite{KlafterRMP}
\begin{equation}
\tilde{\rho}_{r=0}(k,s) = \frac{1+s\tau}{s(1+s\tau)+\tau v^2 k^2}.
\label{eq:fl-diff-lw-no-reset}
\end{equation}
The stationary distribution in the presence of stochastic resetting with instantaneous returns can then be derived, e.g., by use of \eref{renewal-inst}, and one finds \cite{Satya-RT,telegraphic} 
\begin{equation}
\rho_{\infty}(x) = \frac{\alpha_{T}}{2}e^{-\alpha_{T}|x|}~, \end{equation}
with 
\bea
\alpha_{T}=\frac{r}{v}
\sqrt{1 + \frac{1}{r\tau}}~.
\label{Tele_in}
\eea
Comparing \eref{Tele_in} with \eref{DUR_in}, we once again see that all the results obtained for simple diffusion carry through with $\alpha_{T}$ replacing  $\alpha_{0}$. In particular, we note again that the stationary distribution is independent of the return velocity $v_r$ and hence identical to that obtained in the case of instantaneous returns. Our theoretical predictions and associated invariances are corroborated against 
numerical simulations in \fref{fig:hist-tele}. 

\subsection{Fractional diffusion}

As another example of a non-Markovian process, we now consider fractional diffusion which, in the absence of resetting, is described by the fractional Fokker-Planck equation \cite{rw-guide}
\bea
\frac{\partial \rho_{r=0}(x,t)}{\partial t}={}_0 D_t^{1-\gamma} K_{\gamma}~\frac{\partial^2 \rho_{r=0}(x,t)}{\partial x^2}~,
\label{ffpe}
\eea
with $0<\gamma<1$, ${}_0 D_t^{1-\gamma}$ standing for the  Riemann-Liouville fractional derivative operator, and $K_{\gamma}$ for the generalized diffusion coefficient. We recall that $\gamma=1$ corresponds to simple diffusion with $K_1=D$, but that for $\gamma<1$ the process is non-Markovian and subdiffusive with $\langle x^2(t) \rangle=\frac{2K_{\gamma}}{\Gamma(1+\gamma)}t^{\gamma}$ \cite{rw-guide,ergodicity-ctrw}. 

\begin{figure}[t]
\centering
\includegraphics[width=\linewidth]{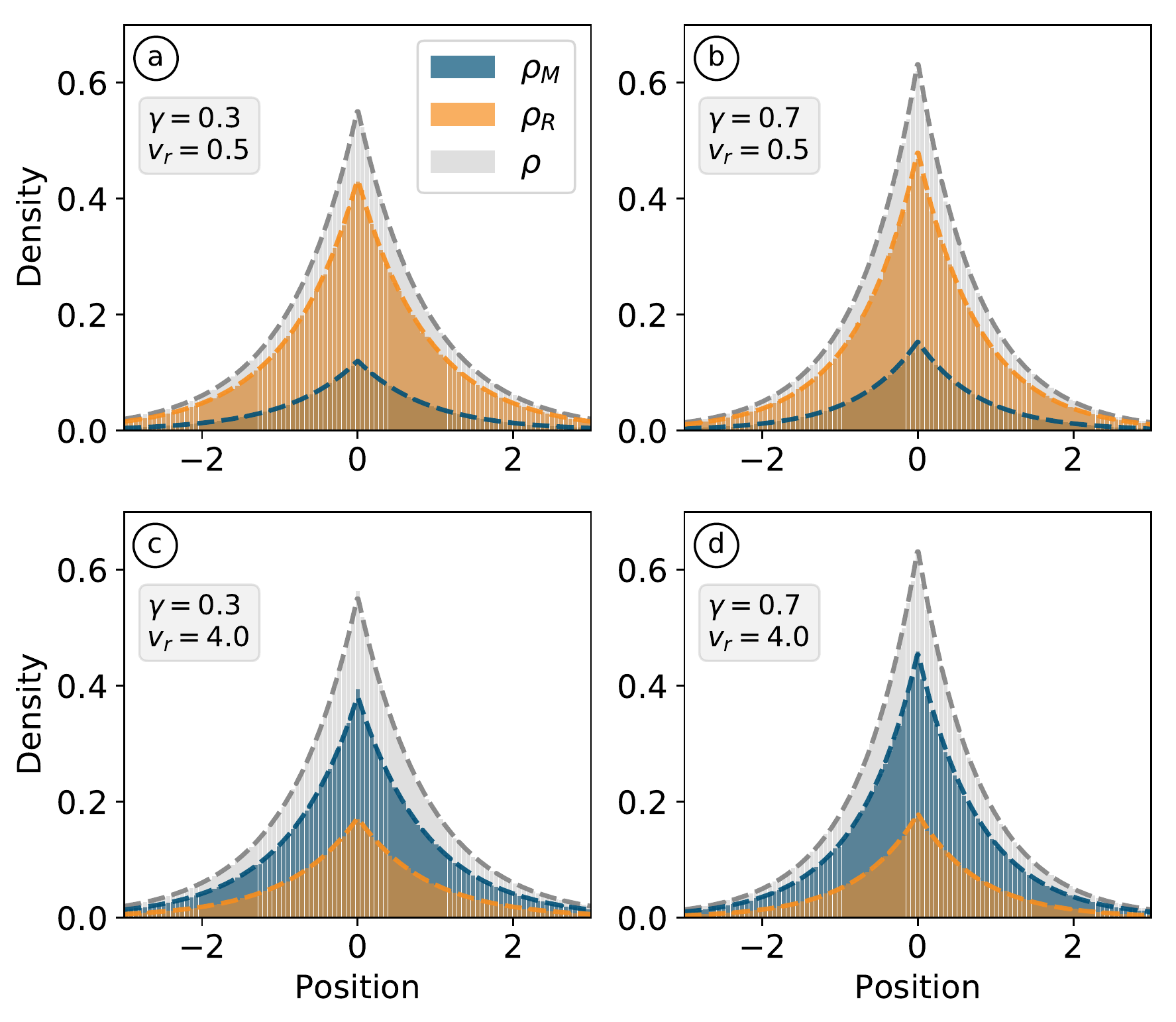}
\caption{
Theory and numerical simulations for fractional diffusion with stochastic resetting and constant speed returns to the origin. Plots are made for four different combinations of the sub-diffusion exponent $\gamma$ and return velocity $v_r$. Histograms represent results obtained via computer simulations, the corresponding dashed lines our theoretical predictions. Details of numerical simulations are identical to \fref{fig:hist-tele}. Excellent agreement with theory is found.
}
\label{fig:hist-subdiff}
\end{figure}

In Laplace space, the solution to \eref{ffpe} is known to be given by \cite{rw-guide}
\bea
\tilde{\rho}_{r=0}(x,s)=\frac{1}{2}~\frac{s^{\gamma/2-1}}{\sqrt{K_\gamma}}~e^{-\sqrt{\frac{s^\gamma}{K_\gamma}}|x|}~.
\eea
Using this form in \eref{renewal-inst}, we obtain the steady-state density of fractional diffusion with stochastic resetting and instantaneous returns \cite{subCTRW}
\bea
\rho_{\infty}(x)=r\tilde{\rho}_{r=0}(x,r)=\frac{\alpha_{\gamma}}{2}e^{-\alpha_{\gamma}|x|}~,
\label{frac_in}
\eea
where $\alpha_\gamma=\sqrt{\frac{r^\gamma}{K_\gamma}}$.
Once again, comparing \eref{frac_in} with \eref{DUR_in}, we see that all the results obtained for simple diffusion carry through with $\alpha_{\gamma}$ replacing $\alpha_{0}$. Our theoretical predictions and associated invariances are corroborated against numerical simulations in \fref{fig:hist-subdiff}.

\begin{figure}[t]
\centering
\includegraphics[width=\linewidth]{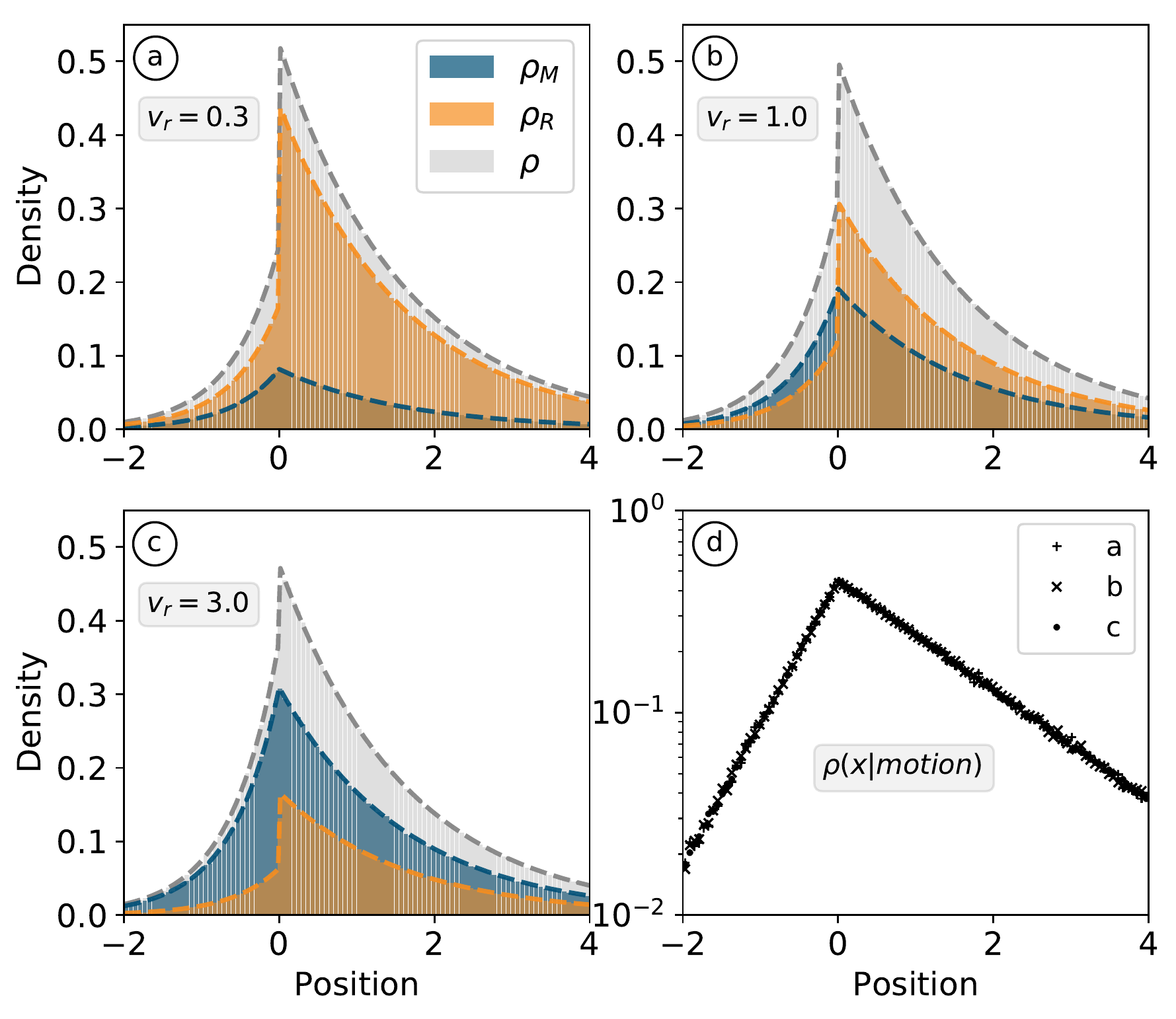}
\caption{Theory and numerical simulations for drift-diffusion with stochastic resetting and constant speed returns to the origin. Plots are made for three different return speeds: (a) $v_r=0.3$; (b) $v_r=1$; and (c) $v_r=3$. Histograms represent results obtained via computer simulations, the corresponding dashed lines our theoretical predictions. In all cases, we took $D=1$, $V=1$, $r=1$, and other details of the numerical simulations are identical to \fref{fig:hist-tele}. Excellent agreement with theory is found. In panel (d), we show numerical results for the conditional steady-state probability density, $\rho(x|motion)$, of finding the particle at $x$ given that it is in the stochastic motion phase. As predicted, this density is seen to be invariant to the return speed.}
\label{fig:hist-drift}
\end{figure}

\subsection{Diffusion with drift}
\label{sec:diff-drift}
In all the examples considered so far the total steady-state density turned out to be completely invariant to the return speed. This invariance could, however, be lost if one starts from an underlying process for which $\rho_{\mathrm{\infty}}(x)$ is not governed by the Laplace distribution. Consider, for example, diffusion with a constant drift velocity $V>0$. The steady-state distribution of this process under stochastic resetting and instantaneous returns is known to be given by \cite{restart_conc2}
\bea
    \rho_{\mathrm{\infty}}(x) = 
    \frac{\alpha_0}{2\sqrt{1 + \lambda^2}} 
    \exp\left[
    - \left(\sqrt{1 + \lambda^2} -\sign(x) \lambda\right)
    \alpha_0 |x|    
    \right],~\hspace{0.6cm}
\eea
where $\alpha_0=\sqrt{\frac{r}{D}}$ and $\lambda = V/(2\sqrt{D r})$. Taking a constant return speed $v_r$, it is easy to compute the probability to be in the drift-diffusion phase using \eref{pa-exact} and we find
\begin{equation}
    p_M = \left[ 1 + \frac{r}{\alpha_0 v_r}\frac{1 + 2\lambda^2}{\sqrt{1+\lambda^2}} \right]^{-1}.
\end{equation}

With $p_M$ and $\rho_{\infty}(x)$ at hand, one can immediately write  $\rho_M(x)=p_M\rho_{\infty}(x)$ for the density in the drift-diffusion phase, and use \eref{eq:passive-from-active-main} to obtain 
\bea
    \rho_R(x) = \frac{r p_M}{2 v_r}\left(
    1 + \frac{\lambda\sign(x)}{\sqrt{1+\lambda^2}}
    \right)
  e^{- \left(\sqrt{1 + \lambda^2} -\sign(x) \lambda\right)
    \alpha_0 |x| 
    }~,\hspace{0.5cm}
\eea
for the density in the return phase. Finally, by summing over the densities in the return and drift-diffusion  phases, one obtains 
\bea
\rho(x)=\left[ 1+\frac{r}{v_r\alpha_0} \left( \sqrt{1+\lambda^2}+\lambda \sgn{x} \right)  \right]p_M\rho_{\infty}(x)~,
\eea
for the total density, which unlike previous examples has an explicit dependence on the return velocity $v_r$. Note, however, that since  $\rho_M(x)=p_M\rho_{\infty}(x)$ we still have $\rho(x|motion)=\rho_{\infty}(x)$. Indeed, and as discussed above, the conditional steady-state density of finding a particle at $x$ given that it is in the stochastic motion phase is an invariant of the return protocol irrespective of the underlying stochastic process which governs motion. Our results for drift-diffusion are corroborated against numerical simulations in Fig. \ref{fig:hist-drift}. 

\subsection{Space dependent return speeds}
\label{sec:vr-depepnds-on-x}

\begin{figure}[t]
\centering
\includegraphics[width=\linewidth]{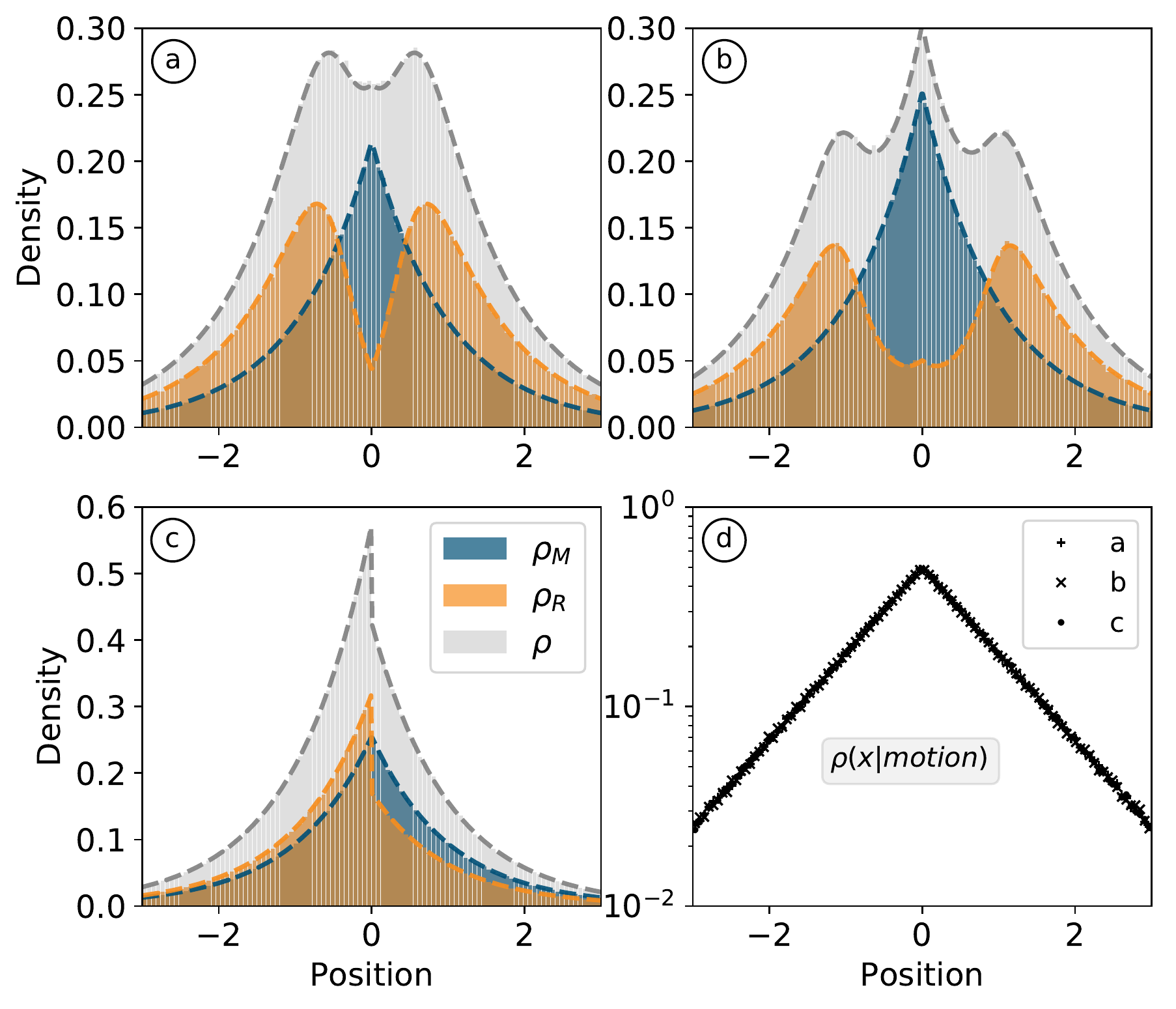}
\caption{Theory and numerical simulations for diffusion with space dependent return speeds: (a) $v_r(x) = 0.5 + 4.5 \exp{(-5 |x|)}$; (b) $v_r(x) = 0.5 + 4.5 \exp{(-3 x^2)}$; and (c) $v_r(x) = 1.1 + 0.4\sign(x)$. In all cases, we took $D=1$, $r=1$, and other details of the numerical simulations are identical to \fref{fig:hist-tele}. Histograms represent results obtained via computer simulations, the corresponding dashed lines our theoretical predictions. Note again, the invariance of $\rho(x|motion)$ in panel (d).}
\label{fig:hist-vr}
\end{figure}

Invariance of the total density can also be broken when the return speed depends explicitly on space. To demonstrate this, we consider simple diffusion once again but now with a return speed, $v_r(x)$, that has some space dependence. Equations (\ref{eq:active-fast-same-main-2}) and (\ref{eq:passive-from-active-main}) then give
\bea
\rho_M(x)&=&\frac{\alpha_0 p_M}{2} e^{-\alpha_0|x|} \\
\label{eq:diff-vx-m}
\rho_R(x)&=&\frac{r p_M}{2 v_r(x)} e^{-\alpha_0|x|}, 
\label{eq:diff-vx-r}
\eea
where $p_M$ is calculated using \eref{pa-exact}. Once again, we observe that since $\rho_M(x)=p_M\rho_{\infty}(x)$, $\rho(x|motion)=\rho_{\infty}(x)$ which does not depend on the return speed despite its space dependence. Note, however, that $\rho(x)$ and $\rho_R(x)$ explicitly depend on the return speed and may exhibit rather non-trial features. For example, a unimodal $v_r(x)$ centered at the origin may lead to bi- or even tri-modal densities [Fig.~\ref{fig:hist-vr}].

\section{Conclusion}
In this paper, we studied the steady-state of a particle undergoing stochastic motion with resetting. We considered a situation where upon resetting the particle does not return to the origin immediately but rather via a prescribed return  protocol, e.g., at a constant speed. We developed a simple, and completely generic, recipe for the computation of the steady-state distributions which govern the stochastic motion and return phases in our model; and showed that the steady-state distribution characterizing the process as a whole follows immediately. We demonstrated the power of our approach on several case studies that illustrated the application of our three step algorithm to find the steady-state distribution:  
\begin{itemize}
    
    \item In the first step, one needs to obtain  $\rho_{\infty}(x)$---the steady-state distribution of the process with stochastic resetting and instantaneous returns to the origin (infinite return speed). This step is now considered common practice, and can e.g., be carried out using a  renewal approach [see \eref{renewal-inst}], or directly by solving for the steady-state of the dynamical equation that describes the process with resetting and instantaneous returns [e.g., \eref{inst-propagator} when it applies].   
    
    \item In the second step, one plugs  $\rho_{\infty}(x)$ into \eref{taux} and utilizes \eref{pa-exact} to compute the steady-state probability, $p_M$, to find the particle in the stochastic motion phase. The steady-state probability to find the particle in the return phase is given by the complimentary probability $p_R=1-p_M$. 
    
    \item In the last step, one plugs  $\rho_{\infty}(x)$ and $p_M$ into Eqs. (\ref{eq:active-fast-same-main-2}) and (\ref{eq:passive-from-active-main}) to obtain the steady-state densities $\rho_{M}(x)$ and $\rho_{R}(x)$ that respectively govern the stochastic motion phase and return phase in our model. The steady-state density which governs the process as a whole is then given by $\rho(x)=\rho_{M}(x)+\rho_{R}(x)$ 
\end{itemize}
The three-step algorithm prescribed above gives a systematic recipe to compute the steady-state of a stochastic process with resetting and space-time coupled returns to the origin. It also reveals two central invariants that are associated with such processes. 

The first invariance is extremely general. It states that: the steady-state density $\rho_{M}(x)$ which governs the stochastic motion phase is nothing but a scaled version of $\rho_{\infty}(x)$---the steady-state density obtained when returns are  instantaneous [see \eref{eq:active-fast-same-main-2}]. This means that the shape of $\rho_{M}(x)$ is invariant to the details of the protocol prescribing how the particle returns to the origin which another way of saying that the conditional steady-state density of finding a particle at $x$ given that it is in the stochastic motion phase is an invariant of the return protocol. As demonstrated above, this invariance result holds for Markovian and non-Markovian processes alike. 

    The second invariance is less general but much more striking. It states that when the return speed is constant, there is a wide class of processes whose steady-state density $\rho(x)$ is completely invariant to whether returns are slow or fast which in particular means that $\rho(x)=\rho_{\infty}(x)$. Specifically, we find that this invariance holds whenever the steady-state density of a process under stochastic resetting and instantaneous returns follows the Laplace distribution: $\rho_{\infty}(x)=\frac{1}{2} \lambda e^{-\lambda|x|}$ with $\lambda>0$. Simple diffusion, diffusion in a $V$-shaped potential, stochastic telegraph motion, and fractional diffusion, are just a few processes that fall into this invariance class. 

The results presented in this work complement those recently presented in \cite{HRS,invariance} and significantly extend our knowledge on motion with stochastic resetting. All current formulations of such motion suffer from the same problem: when considering resetting they neglect the inherent spatio-temporal coupling that governs motion in our world. This crippling situation is in many ways similar to that which hindered the acceptance of the continuous time random walk (CTRW) model \cite{CTRW1,CTRW2,CTRW3,CTRW4} before the development of space-time coupled CTRWs \cite{SPC-CTRW} and L\'evy walks \cite{Walks-CTRW1,Walks-CTRW2,Walks-CTRW3,Walks-CTRW4}. These introduced explicit correlations between time and distance traveled and cured many illnesses of the original CTRW. The space-time coupled version of resetting that was considered herein, and in \cite{HRS,invariance}, is expected to do the same for models of motion with stochastic resetting.    

\section{Author Contributions}
Arnab Pal and \L{}ukasz Ku\'smierz have contributed equally to this work.
\section{Acknowledgements}
All authors would like to acknowledge Tam\'{a}s Kiss, Sergey Denisov and Eli Barkai, organizers of the 672. WE-Heraeus Seminar: ``Search and Problem Solving by Random Walks'', as discussions that led to this work began there. Shlomi Reuveni would like to deeply acknowledge Sidney Redner for a series of joint discussions which led to this work. Shlomi Reuveni acknowledges support from the Azrieli Foundation and from the Raymond and Beverly Sackler Center for Computational Molecular and Materials Science at Tel Aviv University. Arnab Pal acknowledges support from the Raymond and Beverly Sackler Post-Doctoral Scholarship at Tel-Aviv University.

\end{document}